\documentclass[letter,longauth]{aa} 
\usepackage{txfonts} 
\usepackage{graphicx}   
\usepackage{natbib}   
\usepackage{color}   
\usepackage{longtable} 
\bibpunct{(}{)}{;}{a}{}{,} 

\newcommand{\CII}{[\ion{C}{ii}]}

\newcommand{\OI}{[\ion{O}{i}]}

\newcommand{\HII}{\ion{H}{ii}}   
\newcommand{\HI}{\ion{H}{i}}   
   
\newcommand{\NII}{[\ion{N}{ii}]}

\newcommand{\msol}{M$_{\odot}$}   

\newcommand{\herschel}{{\it{Herschel}}}
\newcommand{\spitzer}{{\it{Spitzer}}}

\begin{document}   

\title{PACS and SPIRE photometer maps of M33: First results of the
  \herschel\, M33 extended survey (HERM33ES)\thanks{\herschel\, is an ESA
    space observatory with science instruments provided by
    European-led Principal Investigator consortia and with important
    participation from NASA.} }

   \author{  
     C.\,Kramer\inst{1} \and
     C.\,Buchbender\inst{1}  \and
     E.M.\,Xilouris\inst{2} \and
     M.\,Boquien\inst{3} \and
     J.\,Braine\inst{4} \and 
     D.\,Calzetti\inst{3} \and 
     S.\,Lord\inst{5} \and 
     B.\,Mookerjea\inst{6} \and
     G.\,Quintana-Lacaci\inst{1}  \and
     M.\,Rela\~{n}o\inst{7} \and
     G.\,Stacey\inst{8} \and
     F.S.\,Tabatabaei\inst{9} \and
     S.\,Verley\inst{10} \and
%
     S.\,Aalto\inst{11} \and
     S.\,Akras\inst{2} \and
     M.\,Albrecht\inst{12} \and
     S.\,Anderl\inst{12} \and
     R.\,Beck\inst{9} \and
     F.\,Bertoldi\inst{12} \and
     F.\,Combes\inst{13} \and
     M.\,Dumke\inst{14} \and
     S.\,Garcia-Burillo\inst{15} \and
     M.\,Gonzalez\inst{1}  \and
     P.\,Gratier\inst{4} \and
     R.\,G\"usten\inst{9} \and
     C.\,Henkel\inst{9} \and 
     F.P.\,Israel\inst{16} \and 
     B.\,Koribalski\inst{17} \and 
     A.\,Lundgren\inst{14} \and
     J.\,Martin-Pintado\inst{18} \and
     M.\,R\"ollig\inst{19} \and
     E.\,Rosolowsky\inst{20} \and
     K.F.\,Schuster\inst{21} \and
     K.\,Sheth\inst{22} \and
     A.\,Sievers\inst{1}  \and
     J.\,Stutzki\inst{19} \and
     R.P.J.\,Tilanus\inst{23} \and
     F.\,van der Tak\inst{24} \and
     P.\,van der Werf\inst{16} \and
     M.C.\,Wiedner\inst{13} 
          }   

   \institute{   
     Instituto Radioastronom\'{i}a Milim\'{e}trica, 
     Av. Divina Pastora 7, Nucleo Central, E-18012 Granada, Spain
     \and 
     Institute of Astronomy and Astrophysics, National
     Observatory of Athens, P. Penteli, 15236 Athens, Greece
     \and 
     Department of Astronomy, University of Massachusetts, Amherst, 
     MA 01003, USA 
     \and 
     Laboratoire d'Astrophysique de Bordeaux, Universit\'{e} Bordeaux 1, 
     Observatoire de Bordeaux, OASU, UMR 5804, CNRS/INSU, B.P. 89, 
     Floirac F-33270
     \and 
     IPAC, MS 100-22
     California Institute of Technology, Pasadena, CA 91125, USA
     \and 
     Department of Astronomy \& Astrophysics, 
     Tata Institute of Fundamental Research, 
     Homi Bhabha Road, Mumbai 400005, India
     \and 
     Institute of Astronomy, University of Cambridge, Madingley Road, 
     Cambridge CB3 0HA, England
     \and 
     Department of Astronomy, Cornell University, Ithaca, NY 14853, USA
     \and 
     Max Planck Institut f\"ur Radioastronomie, 
     Auf dem H\"ugel 69, D-53121 Bonn, Germany
     \and 
     Dept. F\'{i}sica Te\'{o}rica y del Cosmos, Universidad de
     Granada, Spain 
     \and 
     Department of Radio and Space Science,
     Onsala Observatory, Chalmers University of Technology, 
     S-43992 Onsala, Sweden
     \and 
     Argelander Institut f\"ur Astronomie. Auf dem H\"ugel 71, 
     D-53121 Bonn, Germany
     \and 
     Observatoire de Paris, LERMA, CNRS, 61 Av. de l'Observatoire, 
     75014 Paris, France
     \and 
     ESO, Casilla 19001, Santiago 19, Chile
     \and 
     Observatorio Astron\'{o}mico Nacional (OAN) - Observatorio de Madrid, 
     Alfonso XII 3, 28014 Madrid, Spain
     \and 
     Leiden Observatory, Leiden University, PO Box 9513,  
     NL 2300 RA Leiden, The Netherlands
     \and 
     ATNF, CSIRO, PO Box 76, Epping, 
     NSW 1710, Australia
     \and 
     Centro de Astrobiolog\'ia (INTA-CSIC),
     Ctra de Torrej\'on a Ajalvir, km 4, 28850 Torrej\'on de Ardoz,
     Madrid, Spain 
     \and 
     KOSMA, I. Physikalisches Institut, Universit\"at zu K\"oln,   
     Z\"ulpicher Stra\ss{}e 77, D-50937 K\"oln, Germany    
     \and 
     University of British Columbia Okanagan, 3333 University Way, 
     Kelowna, BC V1V 1V7, Canada
     \and 
     IRAM, 300 rue de la Piscine, 38406 St. Martin d'H\`{e}res, France
     \and 
     California Institute of Technology, MC 105-24, 1200 East 
     California Boulevard, Pasadena, CA 91125, USA
     \and 
     JAC, 660 North A'ohoku Place, University Park, Hilo, HI 96720,
     USA \and 
     SRON Netherlands Institute for Space Research, Landleven 12, 9747
     AD Groningen, The Netherlands }
   
   \offprints{C.\,Kramer, \email{kramer@iram.es}}   
   \date{For the \herschel\, Special Issue from the HERM33ES Key Programme consortium}
      
   \abstract
   {Within the framework of the HERM33ES key project, we are studying
     the star forming interstellar medium in the nearby, metal-poor
     spiral galaxy M33, exploiting the high resolution and sensitivity
     of \herschel.  }
   {We use PACS and SPIRE maps at 100, 160, 250, 350, and 500\,$\mu$m
     wavelength, to study the variation of the spectral energy
     distributions (SEDs) with galacto-centric distance. }
   {Detailed SED modeling is performed using azimuthally averaged
     fluxes in elliptical rings of 2\,kpc width, out to 8\,kpc
     galacto-centric distance. Simple isothermal and two-component
     grey body models, with fixed dust emissivity index, are fitted to
     the SEDs between 24\,$\mu$m and 500$\,\mu$m using also
     MIPS/\spitzer\, data, to derive first estimates of the dust physical
     conditions. }
   {The far-infrared and submillimeter maps reveal the branched,
     knotted spiral structure of M33. An underlying diffuse disk is
     seen in all SPIRE maps (250--500\,$\mu$m). Two component fits to
     the SEDs agree better than isothermal models with the observed,
     total and radially averaged flux densities.  The two component
     model, with $\beta$ fixed at 1.5, best fits the global and the
     radial SEDs. The cold dust component clearly dominates; the
     relative mass of the warm component is less than 0.3\% for all
     the fits. The temperature of the warm component is not well
     constrained and is found to be about 60\,K$\pm$10\,K. The
     temperature of the cold component drops significantly from
     $\sim24$\,K in the inner 2\,kpc radius to 13\,K beyond 6\,kpc
     radial distance, for the best fitting model. The gas-to-dust
     ratio for $\beta=1.5$, averaged over the galaxy, is higher than
     the solar value by a factor of 1.5 and is roughly in agreement
     with the subsolar metallicity of M33. }
   {}
   \authorrunning{Kramer et al.}
   \maketitle   

\section{Introduction} 

In the local universe, most of the observable matter is contained in
stellar objects that shape the morphology and dynamics of their ``parent"
galaxy. In view of the dominance of stellar mass, a better
understanding of star formation and its consequences is mandatory.
%
%
There exists a large number of high spatial resolution studies related
to individual star forming regions of the Milky Way, as well as of low
linear resolution studies of external galaxies. For a comprehensive
view onto the physical and chemical processes driving star formation
and galactic evolution it is, however, essential to combine local
conditions affecting individual star formation with properties only
becoming apparent on global scales.



At a distance of 840\,kpc \citep{freedman1991}, M33 is the only
nearby, gas rich disk galaxy that allows a coherent survey at high
spatial resolution. It does not suffer from any distance ambiguity, as
studies of the Milky Way do, and it is not as inclined as the
Andromeda galaxy. M33 is a regular, relatively unperturbed disk
galaxy, as opposed to the nearer Magellanic Clouds, which are highly
disturbed irregular dwarf galaxies.

M33 is among the best studied galaxies; it has been observed
extensively at radio, millimeter, far-infrared (FIR), optical, and
X-ray wavelengths, ensuring a readily accessible multi-wavelength
database. These data trace the various phases of the interstellar
medium (ISM), the hot and diffuse, the warm and atomic, as well as the
cold, dense, star forming phases, in addition to the stellar
component.  However, submillimeter and far-infrared data at high
angular and spectral resolutions have been missing so far.


In the framework of the open time key project ``\herschel\, M33
extended survey ({\tt HERM33ES})'', we use all three instruments
onboard the ESA \herschel\, Space Observatory \citep{pilbratt2010} to
study the dusty and gaseous ISM in M33.
%
%
One focus of {\tt HERM33ES} is on maps of the FIR continuum observed
with PACS \citep{poglitsch2010} and SPIRE \citep{griffin2010},
covering the entire galaxy.  A second focus lies on observing
diagnostic FIR and submillimeter cooling lines \CII, \OI, \NII, and
H$_2$O, toward a $2'\times40'$ strip along the major axis with PACS
and HIFI \citep{degraauw2010}.


\begin{figure}[t]   
  \centering   
  \includegraphics[width=7cm]{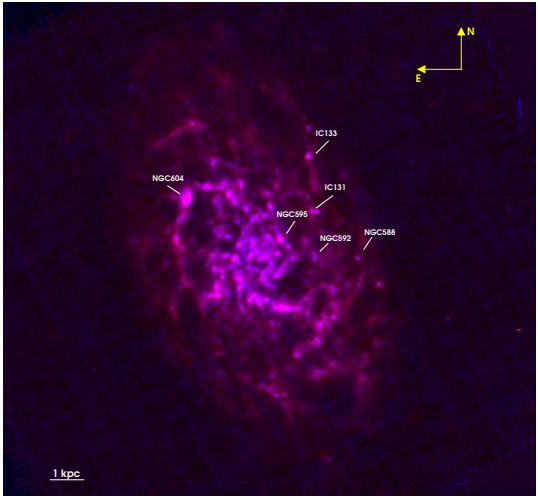} 
  \caption{A composite 500\,$\mu$m (red) and 160$\,\mu$m (blue) map of
    M33. The most extended emission is traced by the longest
    wavelength map revealing the presence of the cold dust in the
    outskirts of the galaxy.  The shorter wavelengths mostly trace the
    branched spiral structure as well as distinct warm \HII\ regions
    and star-forming complexes (such as the ones labeled).}
\label{fig-maps}   
\end{figure}

In this first {\tt HERM33ES} paper, we use continuum maps covering the
full extent of M33, at 100, 160, 250, 350, and $500\,\mu$m. These data
are an improvement over previous data sets of M33, obtained with ISO
and \spitzer\, \citep{hippelein2003,hinz2004,tabatabaei2007-466}, in
terms of wavelength coverage and angular resolution.
%

The total bolometric luminosity of normal galaxies is only about a
factor of 2 larger than the total IR continuum emission
\citep{hauser-dwek2001}, which in turn accounts for more than
$\sim98$\% of the emission of the ISM (dust$+$gas)
\citep[e.g.][]{malhotra2001,dale2001}.
%
%
Massive star formation heats the dust mainly via its far-ultraviolet
(FUV) photons and the absorbed energy is then reradiated in the IR.
FIR continuum fluxes are therefore often used as a measure of the
interstellar radiation field (ISRF) \citep[e.g.][]{kramer2005} and the
star formation rate (SFR) \citep[e.g.][]{schuster2007}.  However, a
number of authors have suggested that half of the FIR emission or more
is due to dust heated by a diffuse ISRF, and not directly linked to
massive star formation \citep{israel1996, verley2009}.

Another disputed topic is the evidence for a massive, cold dust
component in galaxies. The SCUBA Local Universe Galaxy Survey
\citep{dunne2001} identified a cold dust component at an average
temperature of 21\,K. A number of studies of the millimeter continuum
emission of galaxies found indications for even lower temperatures
\citep{misiriotis2006,weiss2008,liu2010}.
%
%
In order to estimate the amount of dust at temperatures below about
20\,K, and to improve our understanding of the physical conditions of
the big grains, well calibrated observations longward of
$\sim150\,\mu$m wavelength are needed.




\section{Observations}

M33 was mapped with PACS \& SPIRE in parallel mode in two orthogonal
directions, in 6.3\,hours on January 7, 2010. Observations were
executed with slow scan speed of $20''$/sec, covering a region of
about $70'\times70'$. Data were taken simultaneously with the PACS
green and red channel, centered on 100 and 160$\,\mu$m. SPIRE
observations were taken simultaneously at 250, 350, and 500$\,\mu$m.
The PACS and the SPIRE data sets were both reduced using the \herschel\,
Interactive Processing Environment (HIPE) 2.0,
with in-house reduction scripts based on the two standard reduction
pipelines.
%

\subsection{PACS data}

The maps are produced with ``photproject'', the default map maker of
the PACS data processing pipeline, and a two-step masking technique.
First we generate a ``naive'' map, i.e.  not properly taking into
account partial pixel overlaps and geometric deformation of the
bolometer matrix, and build a mask considering that all pixels above a
given threshold do not belong to the sky. Then we use this mask to run
the high-pass filter (HPF) taking into account this map.  The mask
helps to preserve the diffuse component to some extent. With new HIPE
tools becoming available, we will try improving data processing to
fully recover the diffuse emission in the PACS maps (cf.
Fig\,\ref{fig-maps}).  The final map is built using the filtered,
deglitched frames. They have a pixel size of 3.2\arcsec\ at 100~$\mu$m
and 6.4\arcsec\ at 160~$\mu$m.  The spatial resolutions of the PACS
data are $6.7''\times6.9''$ at 100$\,\mu$m and $10.7''\times12.1''$ at
160$\,\mu$m.  The pipeline processed data were divided by 1.29 in the
red band and 1.09 in the green one, as this correction is not yet
implemented in HIPE 2.0.\footnote{PACS photometer - Prime and Parallel
  scan mode release note. V.1.2, 23 February 2010} The rms noise
levels of the PACS maps are 2.6~mJy~pix$^{-2}$ at 100~$\mu$m and
6.9~mJy~pix$^{-2}$ at 160~$\mu$m.  The background of the PACS maps of
M33 shows perpendicular stripes in each scanning direction due to
1/$f$ noise.

\subsection{SPIRE data}

A baseline fitting algorithm \citep{bendo2010} was applied to every
scan of the maps.  Next, a ``naive'' mapping projection was applied to
the data and maps with pixel size of $6'', 10''$, and $14''$ were
created for the 250, 350, and 500$\,\mu$m data, respectively.
Calibration correction factors of 1.02, 1.05, and 0.94 were applied to
the 250, 350, and 500$\,\mu$m maps, as this is not yet implemented in
HIPE 2.0.  The spatial resolutions are $18.7''\times17.5''$,
$26.3''\times23.4''$, and $38.1''\times35.1''$ at 250, 350, and
500$\,\mu$m, respectively.  The calibration accuracy is
$15\%$\footnote{SPIRE Beam Model Release Note V0.1, SPIRE Scan-Map AOT
  and Data Products, Issue 2, 21-Oct-2009}. The rms noise levels of
the SPIRE maps of M33 are 14.1, 9.2, and 8\,mJy/beam, at 250, 350,
500$\,\mu$m.

\begin{figure}[t]   
  \centering   
  \includegraphics[width=8cm,angle=-90]{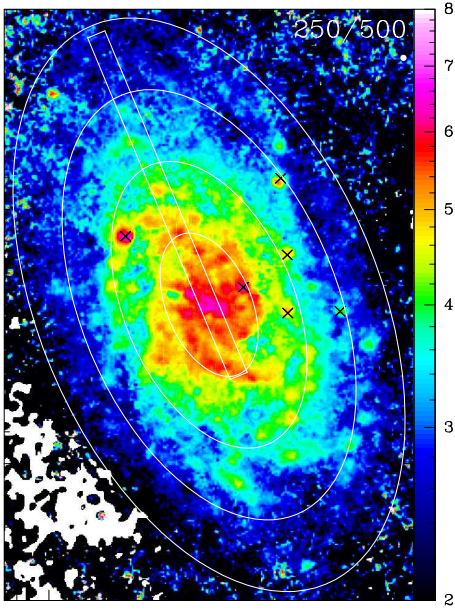} 
  \caption{Ratio map of the $250\,\mu$m over $500\,\mu$m maps, at
    $40''$ resolution. Ellipses denote 2, 4, 6, 8\,kpc galacto-centric
    distances with inclination $56^\circ$ \citep{regan-vogel1994} and
    position angle $22.5^\circ$ \citep{paturel2003}.  The rectangle
    delineates the $2'\times40'$ wide strip along the major axis,
    which will be mapped with HIFI and PACS in spectroscopy mode.
    Crosses mark the \HII\ regions shown in Fig.\,1.}
\label{fig-ratio}   
\end{figure}   

\section{Results}

\subsection{Maps}

Figure\,\ref{fig-maps} shows a composite image of the 160$\,\mu$m,
250$\,\mu$m, and 500$\,\mu$m PACS and SPIRE maps.  All data sets show
the flocculent and knotted spiral arm structure, extending slightly
beyond 4\,kpc radial distance. The PACS $160\,\mu$m map provides the
most detailed view, thanks to its unprecedented linear resolution of
50\,pc, allowing to resolve individual giant molecular clouds (GMCs)
over the entire disk of M33. A large number of distinct sources
delineates the spiral arms.  The properties of these sources are
studied by \citet{verley2010} and \citet{boquien2010}. The SPIRE data
show a faint, diffuse disk, extending out to $\sim7$\,kpc. Outside of
8\,kpc, both maps show some weak emission.

Galactic cirrus is evident only in the outermost part of the galaxy
beyond 6\,kpc radial distance, showing an average contamination of the
order of 2\% which can go up to 8\% at the very faint levels at
500\,$\mu$m. This is still below the 15\% calibration error, which is
the dominant part of the uncertainty.  We did not correct the M33 data
for Galactic Cirrus emission.

\begin{figure}[h]
  \centering   
  \includegraphics[width=9cm]{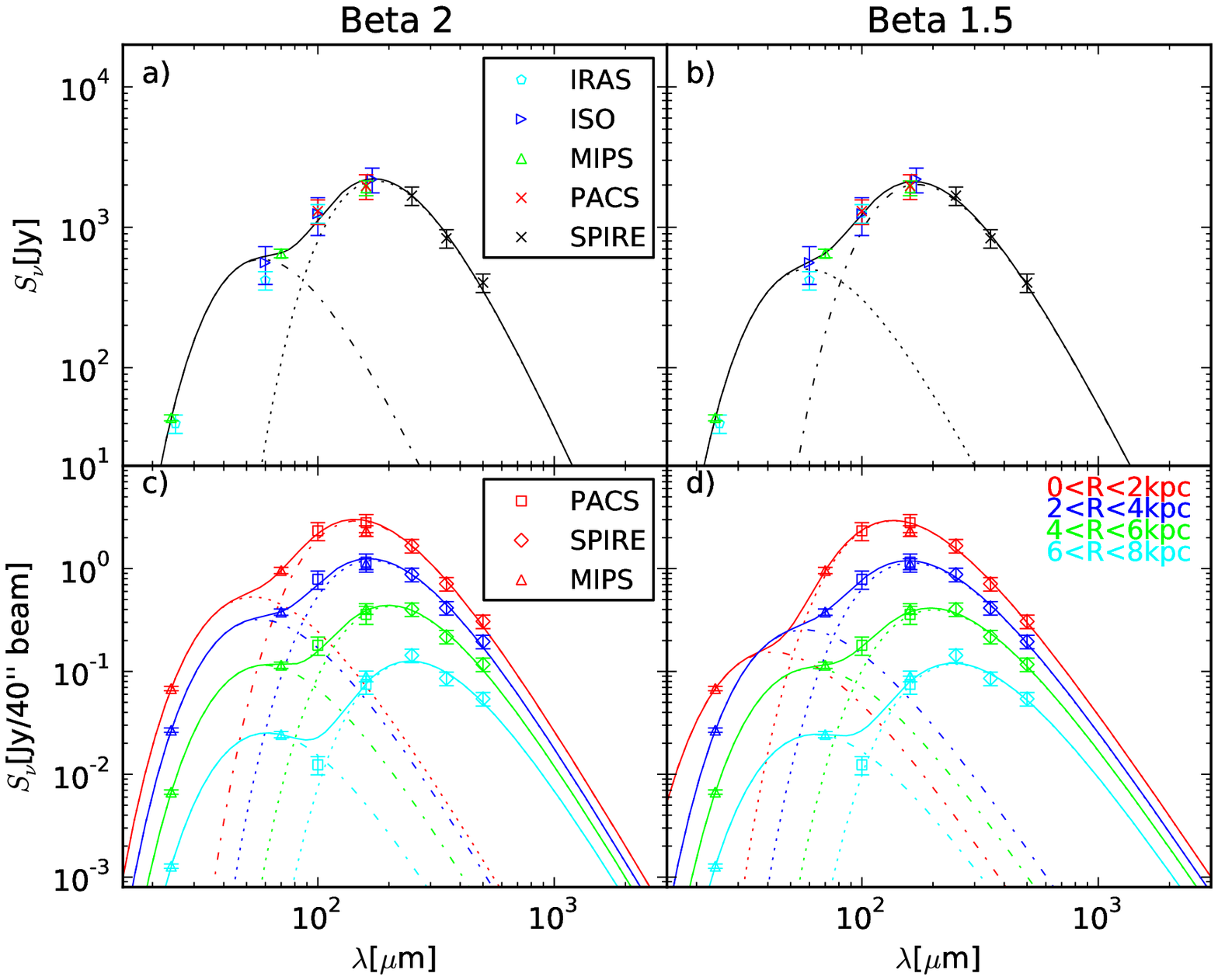} 
  \caption{Spectral energy distributions (SEDs) of M33 at wavelengths
    between 24$\,\mu$m and 500$\,\mu$m. {\bf a, b:} Total integrated
    SED of M33, combining data of PACS \& SPIRE, with data of
    MIPS/\spitzer\, \citep{tabatabaei2007-466}, ISOCAM
    \citep{hippelein2003}, and IRAS \citep{rice1990}. {\bf c, d:}
    Radially averaged SEDs in zones of 2\,kpc width. Here, we show the
    MIPS, PACS, and SPIRE data. {\bf a, c:} Drawn lines show
    two-component grey body model fit results. The 100\,$\mu$m PACS
    flux density, measured in the outermost zone, was not used for the
    fits (see the text).  The dust emissivity index was set to
    $\beta=2$.  {\bf b, d:} Drawn lines show two-component models for
    $\beta=1.5$. {\bf a-d:} All fits are weighted by the assumed
    uncertainties: 7\% for MIPS 24, 70$\,\mu$m
    (\spitzer\, Observers Manual v8.0), 20\% for PACS, 15\% for SPIRE.
    The SEDs have not been de-projected. }
\label{fig-seds}   
\end{figure}

The $S(250\,\mu{\rm m})/S(500\,\mu{\rm m})$ ratio of flux densities
(Fig.\,\ref{fig-ratio}) drops from about $6$ in the inner spiral arms,
to $\sim4$ at $\sim4\,$kpc radius, continuing to less than $\sim3$ at
more than 6\,kpc radial distance.  This drop is also seen in the
radially averaged spectral energy distributions (Fig.\,\ref{fig-seds},
Table\,\ref{tab-seds}).  In addition, the inner spiral arms and a
couple of prominent \HII\ regions (cf.  Fig.\,\ref{fig-maps}),
out to about $5$\,kpc radius, show an enhanced ratio of $\sim6$
relative to the inter-arm ISM, exhibiting a ratio of typically
$\sim4$. This shows that dust is mainly heated by the young massive
stars rather than the general interstellar radiation field in M33.
This is in agreement with a multi-scale study of MIPS data
\citep{tabatabaei2007-466}, where the 160\,$\mu$m emission was found
to be well correlated with H$\alpha$ emission.  .

\begin{center}   
\begin{table}[h]   
  \caption[]{\label{tab-seds} {Results of fits of one and two emission
      components to the measured spectral energy distributions (SEDs) of the
      MIPS, PACS, SPIRE data of M33 shown in Figure\,\ref{fig-seds}. }}
\begin{tabular}{rrrrrrr}   
\hline \hline  
 & Total & (1) & (2) & (3) & (4) \\
\noalign{\smallskip} \hline \noalign{\smallskip}   
 \multicolumn{4}{l}{Isothermal fits} \\
 $T$/[K] & 29 & 25 & 28 & 105 & 37 \\
 $\beta$ & 0.5 & 1.4 & 0.8 & $-1.8$ & $-0.8$ \\
 $\chi^2_{\rm red}$ 
          &   45 &  44 &  45 &  3 & 6 \\  
\noalign{\smallskip} \hline \noalign{\smallskip}   
 \multicolumn{4}{l}{Two-component fits with $\beta=2$} \\ 
 $T_c$/[K] & 16 & 20 & 17 & 14 & 12 \\
 $T_w$/[K] & 50 & 55 & 51 & 49 & 48 \\ 
 $M_c$/[$10^6\,$\msol]        &    8.0  & 1.0  & 2.4  & 3.6  & 4.0 \\
 $M_c/M_w$                    & 1000    & 900  & 900  & 1700 & 5600 \\
 $\chi^2_{\rm red}$ 
                               &    0.41 & 0.23 & 0.33 & 0.42 & 1.8 \\   
 $M_{\rm gas}/M_c$            &    250  & 230  & 180  & 160  & 200 \\
\noalign{\smallskip} \hline \noalign{\smallskip}   
 \multicolumn{4}{l}{Two-component fits with $\beta=1.5$} \\ 
 $T_c$/[K] & 19 & 24 & 20 & 16 & 13 \\
 $T_w$/[K] & 55 & 77 & 57 & 52 & 51 \\
 $M_c$/[$10^{6}$\,\msol] & 10 & 1.2 & 3.0 & 4.6 & 4.9 \\ 
 $M_c/M_w$ & 500 & 3800 & 480 & 730 & 2200 \\
 $\chi^2_{\rm red}$ 
           & 0.14 & 0.10 & 0.12 & 0.20 & 1.8 \\
 $M_{\rm gas}/M_c$ & 200 & 190 & 150 & 120 & 160 \\ 
\noalign{\smallskip} \hline \noalign{\smallskip}   
 \multicolumn{4}{l}{Two-component fits with $\beta=1$} \\ 
 $T_c$/[K] & 23 & 28 & 24 & 19 & 15 \\
 $T_w$/[K] & 62 & - & 67 & 57 & 55 \\
 $M_c$/[$10^6\,$\msol]        & 12   & 1.6  & 3.6  & 5.7  & 5.8 \\
 $M_c/M_w$                    & 300  & -    & 440  & 330  & 870 \\
 $\chi^2_{\rm red}$ 
                              & 0.22 & 0.6 & 0.26 & 0.30 & 2.4 \\
 $M_{\rm gas}/M_c$            & 170  & 140  & 120  & 100   & 140  \\
\noalign{\smallskip} \hline \noalign{\smallskip}   
 \multicolumn{4}{l}{Total gas mass} \\
 $M_{\rm gas}/[10^6\,$\msol] & 2020 & 230 & 440 & 560 & 790 \\ 
\noalign{\smallskip} \hline \noalign{\smallskip}   
\end{tabular}   
\tablefoot{ For the
  two-component fits, the dust emissivity index was kept fixed. $T_c$,
  $T_w$ are the temperatures of the cold and warm component. $M_c$ is
  the total cold dust mass per annulus. $M_c/M_w$ is the dust mass ratio
  of both components. $\chi^2_{\rm red}$ is the
  $\chi^2$ divided by the number of observed parameters minus the number
  of fitted parameters minus 1. The 100\,$\mu$m flux density measured
  in the outermost zone was not used for the fits. The columns give the
  radial annuli: Total: $0<R<8$\,kpc, (1): $0<R<2$\,kpc, (2):
  $2<R<4$\,kpc, (3): $4<R<6$\,kpc, (4): $6<R<8$\,kpc. The last line at
  the bottom of the table gives the total gas masses $M_{\rm
    gas}=1.36\,(M({\rm H}_2)+M($\HI$))$ \citep{gratier2010}.}
\end{table}   
\end{center}   

\subsection{Spectral energy distributions (SEDs)}

Figure\,\ref{fig-seds} shows the total flux densities of M33 and
radially averaged SEDs. The SEDs at different annuli were created by
smoothing all data to a common resolution of $40''$, and averaging the
observed flux densities in radial zones of 2\,kpc width: $r_i\le
R<r_i+2\,{\rm kpc}$ with $r_i=0, 2, 4, 6$\,kpc (cf.
Fig.\,\ref{fig-ratio}). The \herschel\, data agree in general well with
the data from the literature. The MIPS data at 160$\,\mu$m agree
within 20\% with the corresponding PACS data, for all radial zones.
The 100\,$\mu$m PACS flux density, measured in the outermost annulus, is
far below the expected value, indicating that extended, diffuse
emission is at present lost by the data processing. We do not use
these data for the fits.

Figure\,\ref{fig-seds}c,d shows the drop of emission by almost two
orders of magnitude between the center and the outskirts at 8\,kpc
radial distance. One striking feature of the radially averaged SEDs is
the change of the 160/250 PACS/SPIRE flux density ratio (color), which
drops systematically with radial distance, from 1.7 in the inner zone,
to 0.5 in the outer zone.
At the same time, the slope of the SPIRE data turns shallower with
distance, as already seen in Fig.\,\ref{fig-ratio}.
%
%

We fit simple isothermal and two-component grey body models to the
data.  Each component is described by $S_{\nu} = B(\nu,T) \tau_\nu =
B(\nu,T) \kappa_\nu M_d/D^2$, assuming optically thin emission, with
the flux $S_\nu$, the Planck function $B_\nu$, the opacity $\tau_\nu$,
the dust mass $M_d$, the distance $D$, and the dust absorption
coefficient $\kappa_\nu=0.4(\nu/(250\,{\rm GHz}))^\beta$
cm$^2$g$^{-1}$ \citep{kruegel-siebenmorgen1994,kruegel2003}, $\beta$
is the dust emissivity index.  The fit minimizes the function
$\chi^2=\sum ((S_{\nu,{\rm obs}}-S_\nu)/\Delta S_{\nu,{\rm obs}})^2$
using the Levenberg-Marquardt algorithm \citep{bevington1992}, with
the assumed calibration error $\Delta S_{\nu,{\rm obs}}$.  The fits
are conducted at 7 wavelengths using the SPIRE, PACS, and the MIPS
data at 70 and 24\,$\mu$m. The 24\,$\mu$m data helps in constraining
the warm component, though its emission partly stems from
stochastically heated small grains, not only from grains in thermal
equilibrium. To maintain at least two degrees of freedom
\citep{bevington1992} in the 2-component fit, we kept $\beta$ fixed to
values between 1 and 2.  These values are typically found in models
and observations of interstellar dust \citep[see literature compiled
by][]{dunne2001}.

The fits of isothermal models do not reproduce the data well, the
values of the reduced $\chi^2$ are very high. To a large extent, this
is because of the 24\,$\mu$m points, which clearly require a second,
warm dust component.  Two-component grey body models result in a much
better agreement with the data. The best fitting model is the
two-component model with $\beta=1.5$. The $\chi^2_{\rm red}$ values
are better than 0.2 for all annuli out to 6\,kpc, and 1.8 for the
outermost annulus. However, these values are only slightly better or
equal to the $\chi^2_{\rm red}$ values of the two other two-component
models.

We find higher temperatures of the cold component for lower values of
$\beta$, rendering it difficult to determine both parameters at the same
time. This degeneracy between dust temperature and dust emissivity is
a common problem \citep[e.g.][]{kramer2003}. Note, however, that the
total mass of the cold component, is rather well determined. For
$\beta$-values varying between 1 and 2, this mass only varies by
$\sim20$\%.  The masses in each annulus were determined by fitting the
observed SEDs.  As the fitted temperatures are slightly different, the
sum over the four annuli does not exactly agree with the fitted total
cold dust mass in Table\,\ref{tab-seds}.

The cold dust component dominates the mass for the galaxy for all
annuli. Though the warm component is needed to reproduce the data
shortwards of $\sim100\,\mu$m, its relative mass is less than 0.3\%
for all cases.  Therefore, its temperature is not well constrained. It
is found to be about 60\,K$\pm$10\,K.  The temperature of the cold
component is determined to an accuracy of about 3\,K, as estimated
from a Monte Carlo analysis using the observed data with the estimated
accuracies. It drops significantly from $\sim24$\,K in the inner
2\,kpc radius to 13\,K beyond 6\,kpc radial distance, using the best
fitting model.

Table\,\ref{tab-seds} also gives the total gas masses $M_{\rm gas}$ of
the entire galaxy, and of the elliptical annuli.  These are calculated
from \HI\ and CO data presented in \citet{gratier2010}. They assume a
constant CO-to-H$_2$ conversion factor. But note that this X$_{\rm
  CO}$ factor does not strongly affect the total gas masses, as the
\HI\ is dominating.  The gas-to-dust mass ratio for the entire galaxy,
using the best fitting dust model with $\beta=1.5$, is $\sim200$,
about a factor of 1.5 higher than the solar value of 137 \citep[cf.
Table 2 in][]{draine2007}, and a factor of 2 higher than recent dust
models for the Milky Way \citep{weingartner-draine2001,draine2007}.  A
factor of about 2 is expected, as the metallicity is about half solar
\citep{magrini2009}.  The gas-to-dust ratio for $\beta=1.5$ varies
between 200 and 120 in the different annuli. Within our errors this is
consistent with the shallow O/H abundance gradient found by
\citet{magrini2009}.
%
The gas-to-dust ratios found in M33 are similar to the typical values
found in nearby galaxies \citep[e.g.][]{draine2007, bendo2010}.
\citet{braine2010} combine the dust and gas data of M33 to study the
gas-to-dust ratios in more detail and derive dust cross sections.

\begin{acknowledgements} 

  HIPE is a joint development by the \herschel\, Science Ground Segment
  Consortium, consisting of ESA, the NASA \herschel\, Science Center, and
  the HIFI, PACS and SPIRE consortia.
  We would like to thank all those who helped us processing the PACS
  and SPIRE data. In particular we would like to acknowledge support from
  Pierre Royer, Bruno Altieri, Pat Morris, Bidushi Bhattacharya, Marc
  Sauvage, Michael Pohlen, Pierre Chanial, George Bendo. MR
  acknowledges the MC-IEF within the 7$^{\rm th}$ European Community
  Framework Programme.

\end{acknowledgements}   
   
\bibliographystyle{aa} 
\bibliography{aamnem99,14613} 

\end{document}